\documentclass{andp2012}
\usepackage[english]{babel}
\usepackage{color}
\pagespan{}{}
\category{Original Paper}
\subcategory{}
\title{Magnon blockade in a $\mathcal{PT}$-symmetric-like cavity magnomechanical system}
\author[F. Author]{Liang Wang\inst{1}}
\author[F. Author]{Zhi-Xin Yang\inst{1}}
\author[S. Author]{Yu-Mu Liu\inst{2}}
\author[S. Author]{Cheng-Hua Bai\inst{3}}
\author[S. Author]{Dong-Yang Wang\inst{3}}
\author[T. Author]{Shou Zhang\inst{1}
	\footnote{Corresponding author\quad E-mail:~\textsf{szhang@ybu.edu.cn}}}
\author[F. Author]{Hong-Fu Wang\inst{1}
	\footnote{Corresponding author\quad E-mail:~\textsf{hfwang@ybu.edu.cn}}}
\address[1]{Department of Physics, College of Science, Yanbian University, Yanji, Jilin 133002, China}
\address[2]{Department of Physics, College of Science, Northeast Normal University, Changchun, 130024, China}
\address[3]{School of Physics, Harbin Institute of Technology, Harbin, Heilongjiang 150001, China}
\begin{abstract}
 We investigate the magnon blockade effect in a parity-time ($\mathcal{PT}$) symmetric-like three-mode cavity magnomechanical system involving the magnon-photon and magnon-phonon interactions.
 In the broken and unbroken $\mathcal{PT}$-symmetric regions, we respectively calculate the second-order correlation function analytically and numerically and further determine the optimal value of detuning. By adjusting different system parameters, we study the different blockade mechanisms and find that the perfect magnon blockade effect can be observed under the weak parameter mechanism. Our work paves a way to achieve the magnon blockade in experiment.
\end{abstract}
\shortabstract
\begin{document}
\maketitle
\section{\label{sec.1}Introduction}
Single-photon source has attracted extensive attention and has important applications in the field of quantum optics~\cite{PhysRevLett.69.1293,PhysRevA.69.032305,RevModPhys.79.135}. Photon blockade, which describes a phenomenon that the second photon is blockaded once the first photon has been excited because of the nonlinearity of unequal level spacing, can be used to generate the single-photon source because of the antibunching photons statistic characteristics. Generally, conventional photon blockade (CPB)~\cite{PhysRevLett.79.1467} represents off-resonance in the second-photon transition due to the anharmonicity of the eigenenergy spectrum, which has been studied in cavity-QED systems~\cite{Nature43,PhysRevLett.114.233601,PhysRevA.96.011801,PhysRevLett.118.133604}, optomechanical systems~\cite{PhysRevA.96.013861,PhysRevA.99.043818,PhysRevA.92.033806,PhysRevLett.107.063601}, nonlinear optical systems~\cite{PhysRevA.82.053836} and so on~\cite{PhysRevA.87.023822,PhysRevA.89.043818,PhysRevA.87.023809}. Furthermore, unconventional photon blockade (UPB)~\cite{PhysRevA.83.021802}, which is based on 
the destructive quantum interference effect between two different excitation pathways, has also been studied in coupled Kerr-cavity systems~\cite{PhysRevA.96.053810,Science}, double-cavity optomechanical systems~\cite{PhysRevA.87.013839}, and other systems~\cite{PhysRevA.88.033836,PhysRevA.92.023838,PhysRevA.89.031803}. 
In recent years, the investigation of quantum characteristics of the magnon is attracting extensive attention. Magnon blockade, as a pure quantum phenomenon, is one of the most important aspects of exploring quantum propertiess, which arises from the anharmonicity in energy eigenvalues of an magnon mode. It is necessary for the preparation of single magnon sources. There are two general ideas: (i) the conventional magnon blockade is based on the anharmonicity of the eigenenergy spectrum coming from kinds of nonlinearities; (ii) the unconventional magnon blockade is based on the destructive quantum interference between different excitation paths. Also, magnon blockade has been studied in a hybrid ferromagnet-superconductor quantum system\cite{PhysRevB.100.134421}. Many nonlinear optical effects have attracted intense studies and opened up a promising way to study other important optomechanics and magnetomechanics effects based on the intrinsic properties of optical and magnon systems, e.g., optomechanically induced transparency~\cite{PhysRevA.97.013843} or squeezing~\cite{Wollman952}, entanglement~\cite{PhysRevLett.98.030405} and magnon Kerr effect ~\cite{PhysRevB.94.224410}.

The cavity-magnon systems, which consist of a microwave cavity and a yttrium iron garnet (YIG) sphere, have been theoretically proposed and experimentally realized~\cite{PhysRevLett.113.156401,PhysRevLett.111.127003,PhysRevLett.113.227201,PhysRevB.91.214430,PhysRevApplied.2.054002} and have provided a specific platform for investigating quantum coherence. The ferromagnetic resonance mode (Kittel mode)~\cite{PhysRev.73.155} can be strongly coupled to microwave cavity photons due to low damping rate and high spin density of YIG sphere. Besides, the deformation of YIG sphere is influenced by magnon and results in phonon-magnon interaction. Meanwhile, the cavity magnomechanical system~\cite{Zhange1501286,PhysRevResearch.1.033161} offers a great opportunity for coherent quantum information processing~\cite{PhysRevLett.1.241}. Furthermore, various interesting phenomena have been studied in cavity magnomechanical systems, such as magnon-induced transparency~\cite{OE1}, squeezed states of magnons and phonons~\cite{PhysRevA.99.021801}, magnon-photon-phonon entanglement~\cite{PhysRevLett.121.203601}, etc. Compared with radiation force and electrostatic force in different optomechanical systems~\cite{RevModPhys.86.1391,ADP1,PhysRevLett.94.223902,OE,Gigan2006,ADP2,Bagci2014}, magnetostrictive force possesses good tunability, which describes the electromagnetic force density acting on a magnetic medium. On the other hand, although  $\mathcal{PT}$-symmetry system is non-Hermitian, it shows the characteristics of Hermitian system, which can act as a platform for investigating different quantum behaviors and has been explored in various physical systems~\cite{PhysRevLett.80.5243,PhysRevLett.109.033902,PhysRevA.96.043810,PhysRevLett.103.093902,Regensburger2012,Nature1}. 

In this paper, based on the magnon-photon and magnon-phonon interactions, we propose a scheme to investigate magnon blockade in a $\mathcal{PT}$-symmetric-like cavity magnomechanical system. Generally, the $\mathcal{PT}$-symmetric optomechanical system is a two-mode system, while here we consider a three-mode system consisting of microwave cavity mode, magnon mode, and phonon mode. We study the phase transition points at broken and unbroken $\mathcal{PT}$-symmetric regions. To achieve the magnon blockade, we calculate the second-order correlation function analytically and numerically. Particularly, we also analyze the anharmonicity of the eigenenergy spectrum and study the optimal condition of the magnon blockade. Moreover, we show that the magnon blockade effect can be achieved by adjusting the detuning, magnon-photon coupling strength, and the nonlinear parameters in $\mathcal{PT}$-symmetric-like region. Furthermore, we find the phenomenon of broken and unbroken $\mathcal{PT}$-symmetric regions for different detunings on magnon blockade.  

The paper is organized as follows: In Sec.~\ref{sec.2}, we give the physical model and the Hamiltonian of system. In Sec.~\ref{sec.3}, we study the phase transition in the $\mathcal{PT}$-symmetric-like region. In Sec.~\ref{sec.4}, to observe the magnon blockade effect, the second-order correlation function is calculated analytically and numerically. Finally, a conclusion is given in Sec.~\ref{sec.5}.

\begin{figure}
\centering
\includegraphics[width=1.0\columnwidth]{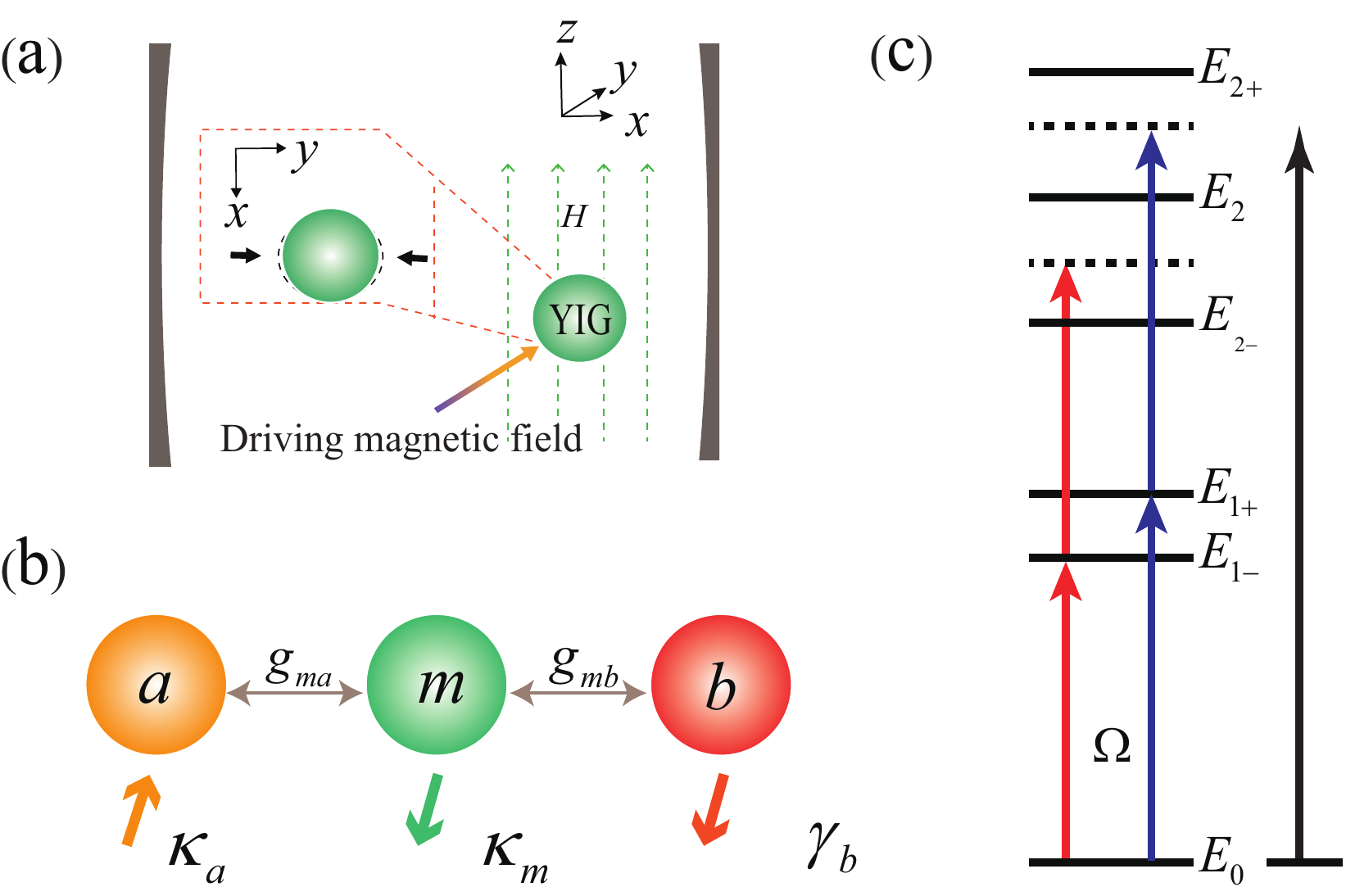}
\caption{(Color online) (a) Schematic diagram of $\mathcal{PT}$-symmetric-like cavity-magnon system, a YIG sphere is placed inside a microwave cavity, the magnetic field and the drive magnetic field  of the cavity mode are in the \textsl{x} and \textsl{y} direction respectively, and the bias magnetic field is applied in the $z$-direction. The ellipse with the black dashed curve (graphs in dotted lines) denotes the deformation caused by the magnetostrictive force. (b) The coupled-harmonic-resonator model, the parameters $\kappa_a$, $\kappa_m$, and $\gamma_b$ denote the decay rates of the microwave cavity mode, magnon mode, and mechanical resonator mode, respectively. (c) The energy-level diagram of the cavity-magnon coupling system.}\label{fig1}	
\end{figure}
\section{\label{sec.2}System and Hamiltonian}
As shown in \textbf{Figure~\ref{fig1}}a, the cavity magnomechanical system under consideration consists of a microwave cavity and a YIG sphere, which involves the magnon-photon and magnon-phonon interactions induced by magnetic dipole and magnetostriction, respectively. The change of the magnetization caused by the excitation of the magnon in the YIG sphere results in the deformation of its geometry, which forms vibrational modes (phonons) of the sphere, and vice versa. In the position of the YIG sphere, the direction of the magnetic field of the cavity mode is along \textsl{x} axis, the external driving magnetic field is along \textsl{y} axis, and an additional bias magnetic field along $z$ axis is placed in the microwave cavity. In such a physical model, the magnetostrictive effect leads to the coupling between the photon and the magnon. The magnon-photon coupling strength can be tuned by the extra magnetic field and the position of the YIG sphere. The magnon in sphere of YIG can achieve strong coupling to the microwave cavity due to its high spin density $(4.2\times10^{21} \mathrm{cm}^{-3})$. For convenience, we present the equivalent model in \textbf{Figure~\ref{fig1}}b. Under the rotating wave approximation, the Hamiltonian of the system reads~\cite{Zhange1501286} $(\hbar=1)$
\begin{eqnarray}\label{eq1}
H&=&\omega_aa^{\dag}a+\omega_bb^{\dag}b+\omega_mm^{\dag}m+ g_{ma}(m^{\dag}a+a^{\dag}m)
\cr\cr &&-g_{mb}(b^\dag+b)m^{\dag}m+K(m^{\dag}m)^2
\cr\cr &&+{\Omega}(m^{\dag}e^{-i{\omega}_lt}+me^{i{\omega}_lt}),
\end{eqnarray}
where $a$ ($a^{\dag}$), $b$ ($b^{\dag}$), and $m$ ($m^{\dag}$)  are the annihilation (creation) operators of the cavity mode, mechanical mode, and magnon mode, respectively, with resonance frequency $\omega_a$, $\omega_b$, and $\omega_m$. $g_{ma}$ and $g_{mb}$ denote the magnon-photon and magnon-phonon coupling strengths, respectively. The single-magnon magnomechanical coupling rate $g_{mb}$ is typically small, but the magnomechanical interaction can be enhanced by driving the magnon mode with a strong microwave field~\cite{PhysRevLett.121.203601}. $K$ is the Kerr nonlinear coefficient and $\Omega$ represents the external driving amplitude acting on the magnon mode. $K(m^{\dag}m)^2$ represents the Kerr effect of magnons owing to the magnetocrystalline anisotropy in the YIG sphere. In the rotating frame with respect to $U=$exp$[-i\omega_{l}t(m^{\dag}m+a^{\dag}a)]$, the transformed Hamiltonian $H_1={U^\dag}HU-i{U^\dag}\dot{U}$ with the form
\begin{eqnarray}\label{eq2}
H_1&=&\Delta_aa^{\dag}a+\omega_bb^{\dag}b+\Delta_mm^{\dag}m+ g_{ma}(m^{\dag}a+a^{\dag}m)\cr\cr
&&-g_{mb}(b^\dag+b)m^{\dag}m+K(m^{\dag}m)^2+{\Omega}(m^{\dag}+m),
\end{eqnarray}
where $\Delta_a=\omega_a-\omega_l$ and $\Delta_m=\omega_m-\omega_l$ are the detunings of the microwave cavity mode and magnon mode, respectively. When a unitary transformation $U=$exp$[g_{mb}/{\omega_{b}}(b^{\dag}-b)]$ is applied to $H_1$ and in the case of $g_{mb}\ll\omega_b$, the Hamiltonian can be obtained as
\begin{eqnarray}\label{eq3}
H_2&=&\Delta_aa^{\dag}a+\omega_bb^{\dag}b+\Delta_mm^{\dag}m+ g_{ma}(m^{\dag}a+a^{\dag}m)\cr\cr&&+N(m^{\dag}m)^2+{\Omega}(m^{\dag}+m),
\end{eqnarray}
where $N=K-g_{mb}^2/\omega_{b}$. As shown in above equation, the system discouples with the dynamics of the mechanical mode, thus we can safely ignore the mechanical mode. Therefore, the Hamiltonian can be further expressed as
\begin{eqnarray}\label{eq4}
H_3&=&\Delta_aa^{\dag}a+\Delta_mm^{\dag}m+g_{ma}(m^{\dag}a+a^{\dag}m)
\cr\cr
&&+N(m^{\dag}m)^2+{\Omega}(m^{\dag}+m).
\end{eqnarray}
Taking into account the microwave cavity gain and magnon mode loss, the Hamiltonian of non-Hermitian system is given by
\begin{eqnarray}\label{eq5}
H_4&=&H_3+i\frac{\kappa_a}{2}a^{\dag}a-i\frac{\kappa_m}{2}m^{\dag}m,
\end{eqnarray}
where $\kappa_a$ is the gain rate of cavity mode and $\kappa_m$ is the decay rate of the magnon mode. The non-Hermitian Hamiltonian is related to $\mathcal{PT}$-symmetry, thus we consider the effcet of phase transition point of non-Hermitian system on magnon blockade which distinguishes the dynamical phenomena of broken $\mathcal{PT}$-symmetric and unbroken $\mathcal{PT}$-symmetric regions.
\section{\label{sec.3}PHASE TRANSITION}
Under the condition of weak nonlinearity, the Hamiltonian in Equation (5) can be further reduced as 
\begin{eqnarray}\label{eq6}
H_5&=&(\Delta_a+i\kappa_a/2)a^{\dag}a+(\Delta_m-i\kappa_m/2)m^{\dag}m
\cr\cr &&+g_{ma}(m^{\dag}a+a^{\dag}m),
\end{eqnarray}
where we have dropped the driving term. To explore the physical process more clearly, we expand the above Hamiltonian by the vector $M=[m, a]^T$ and a $2\times2$ matrix is obtained
\begin{eqnarray}\label{eq7}
H_k=
\begin{bmatrix}
{\Delta_m}-i\kappa_m/2 &  g_{ma} \\
g_{ma} & {\Delta_a}+i\kappa_a/2
\end{bmatrix}.
\end{eqnarray}

\begin{figure}
	\includegraphics[width=0.5\columnwidth]{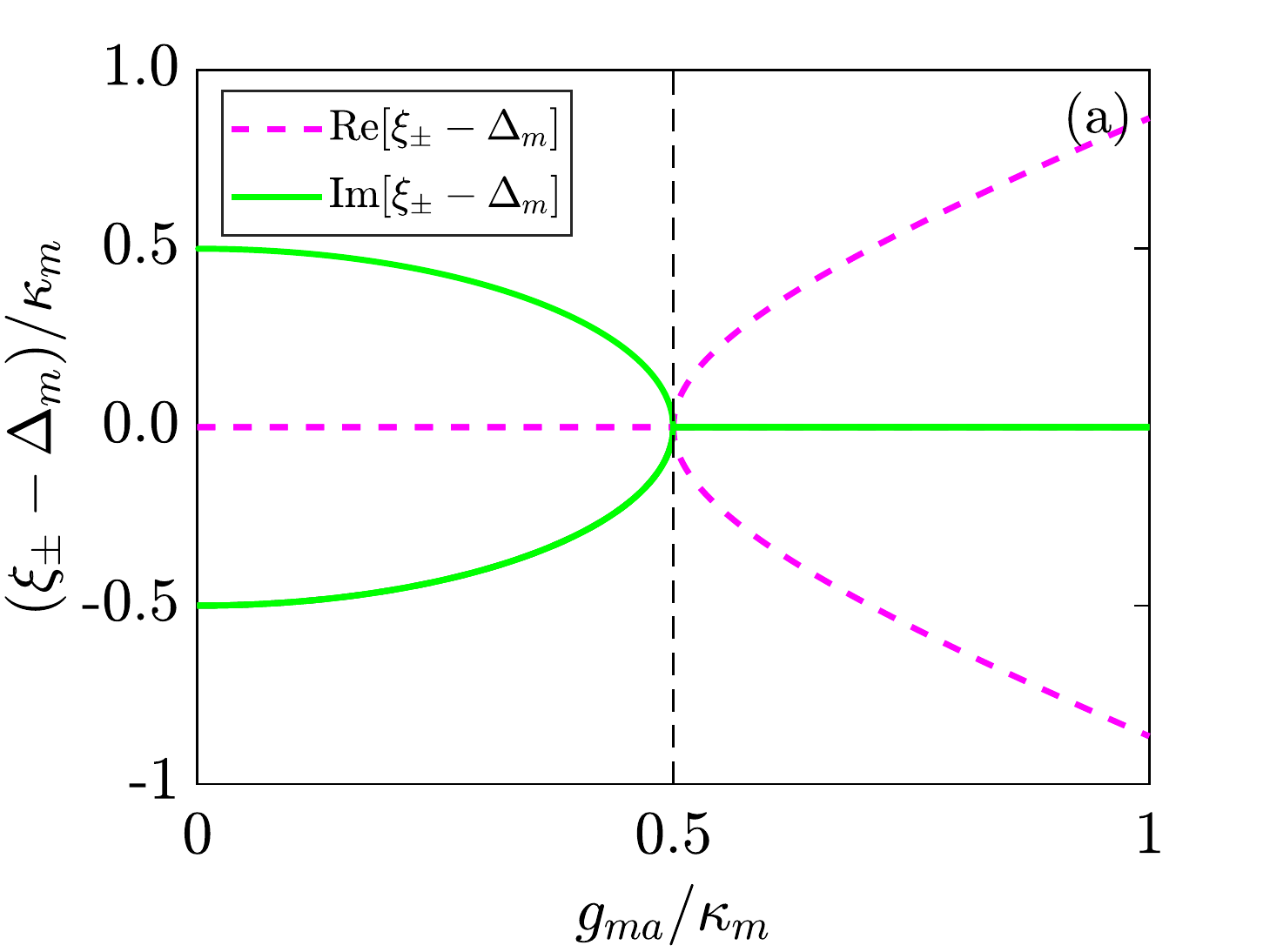}
	\hspace{0.0in}
	\includegraphics[width=0.5\columnwidth]{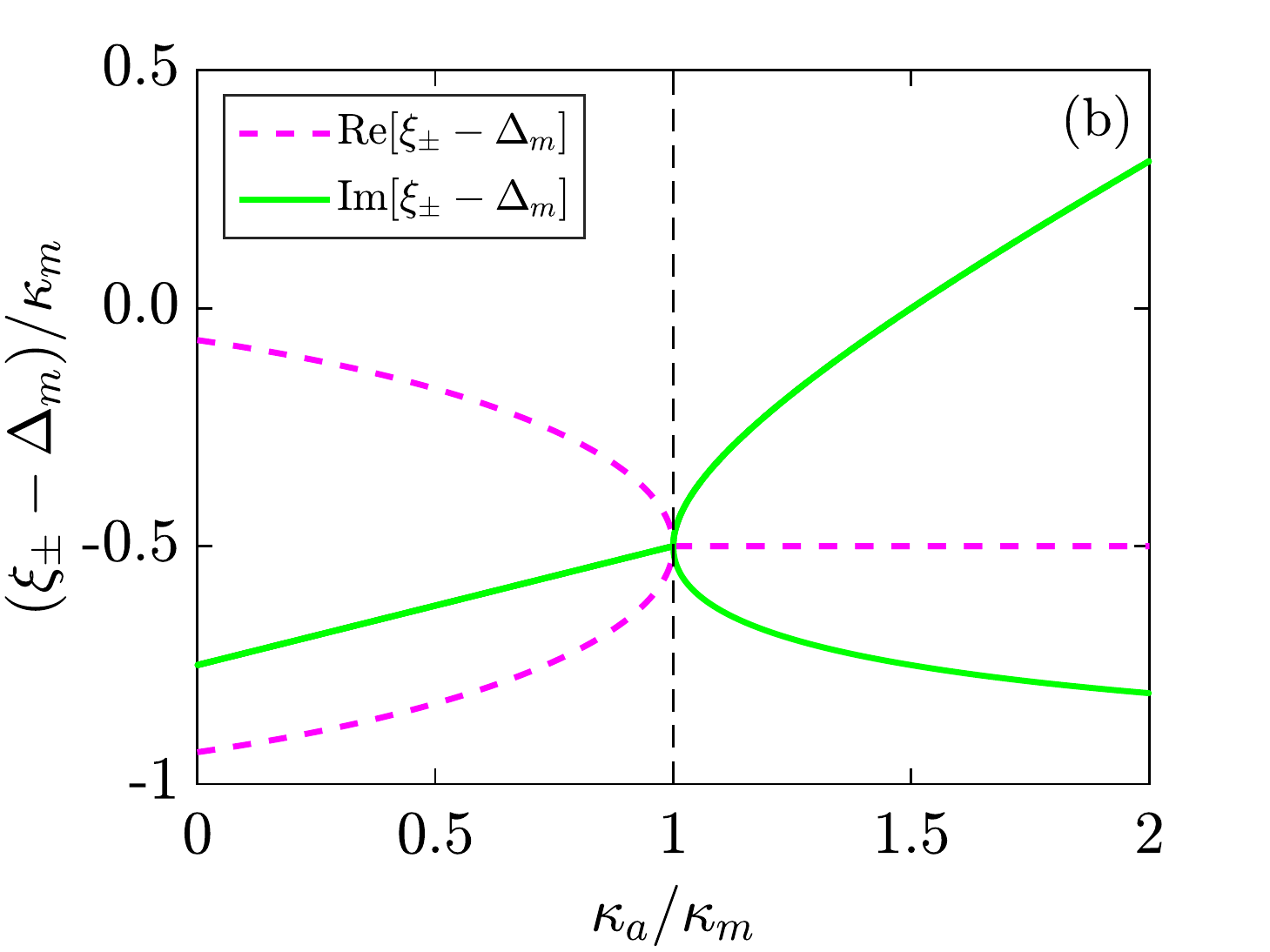}
	\caption{(Color online) The real and imaginary parts of the eigenvalues ($\xi_\pm-\Delta_m$) versus the  magnon-photon strength $g_{ma}$ and gain $\kappa_a$. The pink dashed and green solid lines denote the real and imaginary parts of eigenvalues, respectively. Here we set (a) $\kappa_a/\kappa_m=1$, (b) $g_{ma}/\kappa_m=0.5$.}\label{fig2}
\end{figure}

When $\Delta_a$=$\Delta_m$, the eigenvalues of matrix $H_k$ are $\xi_\pm=\Delta_m- i\lambda/2\pm\sqrt{g_{ma}^2-(\frac{\kappa_a+\kappa_m}{4})^2}$ with $\lambda=(\kappa_a-\kappa_m)/2$. In \textbf{Figure~\ref{fig2}}a, we plot the real and imaginary parts of the eigenvalue ($\xi_\pm-\Delta_m$) versus the magnon-photon strength $g_{ma}$ in a $\mathcal{PT}$-symmetric-like hybrid cavity magnon system. Here we consider three different cases: (i) the magnon-photon coupling strength $g_{ma}$ is less than the effective loss $(\kappa_a+\kappa_m)/4$, i.e., $\kappa_a=\kappa_m$, in which the two modes become degenerate with the opposite imaginary part and the system has a $\mathcal{PT}$-symmetry breaking phase; (ii) the magnon-photon coupling strength $g_{ma}$ is equal to the effective loss $(\kappa_a+\kappa_m)/4$, leading to a phase transition point, which is also called exceptional point; (iii) when the coupling strength $g_{ma}$ is larger than effective loss $(\kappa_a+\kappa_m)/4$, the eigenvalue is nondegenerate but two modes have the same linewidth. When $g_{ma}$ is larger than the critical value, it corresponds to the unbroken $\mathcal{PT}$-symmetric region, vice versa. In figure 2(b), we plot the real and imaginary parts of the eigenvalue $(\xi_\pm-\Delta_m)$ versus the gain rate $\kappa_a$. When the gain and loss are balanced, one can see that the real and imaginary parts are separated and closed, respectively.

\section{\label{sec.4}MAGNON BLOCKADE IN THE $\mathcal{PT}$-SYMMETRIC-LIKE MAGNOMECHANICAL SYSTEM}
In this section, we analytically and numerically calculate the second-order correlation function to study the magnon blockade effect.
\subsection{Analytical solution}
The system we consider is a $\mathcal{PT}$-symmetric-like hybird cavity magnon system, which can be truncated up to 2 in the weak driving limit, the state of the system is thus given by
\begin{eqnarray}\label{eq8}
|\psi(t)\rangle&=& C_{00}(t)|0,0\rangle+C_{10}(t)|1,0\rangle++C_{20}(t)|2,0\rangle\cr\cr
&&+C_{01}(t)|0,1\rangle+C_{11}(t)|1,1\rangle+C_{02}(t)|0,2\rangle.
\end{eqnarray}
where $C_{nm}$ are the amplitudes of the quantum states $|n,m\rangle$. By solving Schr\"{o}dinger equation $i\partial |\psi\rangle/\partial t=H_5|\psi\rangle$, the probability amplitudes can be calculated by the following differential equations
\begin{eqnarray}\label{eq9}
i\frac{\partial C_{00}}{\partial t}&=&EC_{10},\cr\cr
i\frac{\partial C_{10}}{\partial t}&=&EC_{00}+(\Delta_m-i\kappa_m/2-g_{mb}^2/\omega_b+K)C_{10}
\cr\cr &&+g_{ma}C_{01}+\sqrt{2}EC_{20},\cr\cr
i\frac{\partial C_{20}}{\partial t}&=&2(\Delta_m-i\kappa_m/2-2g_{mb}^2/\omega_b+2K)C_{20}
\cr\cr &&+\sqrt{2}EC_{10}+\sqrt{2}g_{ma}C_{11},\cr\cr
i\frac{\partial C_{01}}{\partial t}&=&g_{ma}C_{10}+EC_{11}+(\Delta_m+i\kappa_m/2)C_{01},\cr\cr
i\frac{\partial C_{11}}{\partial t}&=&\sqrt{2}g_{ma}C_{20}+EC_{10}+\sqrt{2}g_{ma}C_{02}
\cr\cr &&+(2\Delta_m+K-g_{mb}^2/\omega_b)C_{11},\cr\cr
i\frac{\partial C_{02}}{\partial t}&=&\sqrt{2}g_{ma}C_{11}+2(\Delta_m+i\kappa_m/2)C_{02}.
\end{eqnarray}
Under the weak driving regime, we consider that $\{C_{02}$, $C_{11}$, $C_{20}\}\ll\{{C_{01}, C_{10}}\}\ll\ C_{00}$. Setting $C_{00}\simeq1$, we can obtain the steady-state solution and the optimal parameters approximately (see Appendix). The value of the second-order correlation function $g_m^{(2)}(0)<1$ corresponds to sub-Possonian statistics of the magnon, then the magnons are antibunched, vice versa. In the following, we investigate the occurrence of magnon blockade by analyzing the magnon statistics, which are characterized by the second-order correlation function in the steady state, given by
\begin{eqnarray}\label{eq10}
g_m^{(2)}(0)=\frac{2|C_{20}|^2}{(|C_{10}|^2+|C_{11}|^2+2|C_{20}|^2)^2} \simeq\frac{2|C_{20}|^2}{|C_{10}|^4}.
\end{eqnarray}
The magnon blockade can be achieved in the case of $|C_{20}|=0$ and the system parameters should satisfy: $\mathrm{\Delta_{opt}}=\Delta_m=g_{mb}^2/2\omega_b-K/2$, which is given in Appendix. 
\subsection{Numerical solution}
The accurate result of magnon blockade can be obtained by numerically calculating quantum master equation, which is given as
\begin{eqnarray}\label{eq11}
\dot{\rho}&=&i[{\rho},H_1]-\frac{\kappa_a}{2}L_a[{\rho}]+\frac{\gamma_b}{2}(n_b^{\mathrm{th}}+1)L_b({\rho})\cr\cr &&+\frac{\gamma_b}{2}n_b^{\mathrm{th}}L_{b^{\dag}}({\rho})+\frac{\kappa_m}{2}(n_m^{\mathrm{th}}+1)L_m({\rho})\cr\cr
&&+\frac{\kappa_m}{2}(n_m^{\mathrm{th}}+1)L_{m^{\dag}}({\rho}),
\end{eqnarray}
where
\begin{eqnarray}\label{eq12}
L_o[{\rho}]=(2o\rho o^{\dag}-o^{\dag}o\rho-\rho o^{\dag} o)
\end{eqnarray}
is the standard Lindblad operator for the arbitrary system operator $o$ and $\gamma_b$ represents 
the dissipation rate of machanical resonator. $n_b^{\mathrm{th}}=\left[\mathrm{exp}(\hbar\omega_b/k_BT)-1\right]^{-1}$ and $n_m^{\mathrm{th}}=\left[\mathrm{exp}(\hbar\omega_m/k_BT)-1\right]^{-1}$ are the thermal excitation numbers of mechanical and magnon modes at temperature $T$, and $k_B$ is the Boltzmann constant. Simply, we assume $n_b^{\mathrm{th}}$=$n_m^{\mathrm{th}}$=$n_{\mathrm{th}}$ in the rest of the paper. We numerically calculate $g_m^{(2)}(0)$ by solving Eq.~(\ref{eq11})
\begin{eqnarray}\label{eq13}
g_m^{(2)}(0)=\frac{\mathrm{Tr}
	({m^{\dag}}^2m^2\rho)}{[\mathrm{Tr}
	(m^{\dag}m\rho)]^2}.
\end{eqnarray}

\begin{figure}
	\centering
	\includegraphics[width=1.0\linewidth]{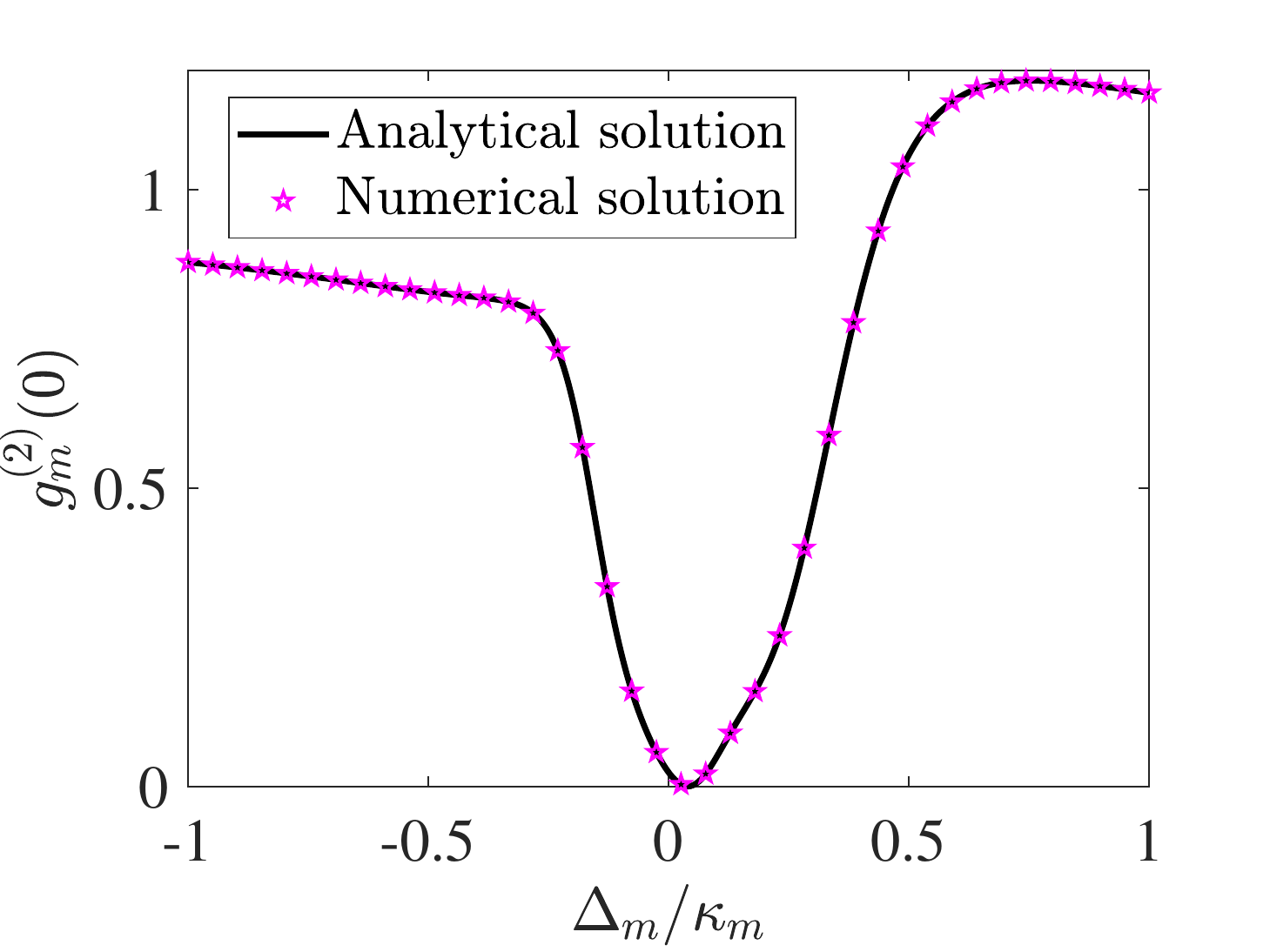}
	\caption{(Color online) The equal-time second-order correlation functions versus the detuning $\Delta_m$. The black solid line represents the analytical solution in Equation (10) and the pink star line represents the numerical solution in Equation (11), respectively. The system parameters are set as $\kappa_m/2\pi$=1 MHz, $\kappa_a=\kappa_m$, $\Delta_a=\Delta_m$, $\omega_m/\gamma_m=10^6$,  $\omega_m/\kappa_m=100$, $K/\kappa_m=0.01$, $g_{ma}/\kappa_m=0.5$, $g_{mb}/\kappa_m=3$, $\Omega/\kappa_m=0.01$, and $n_{\mathrm{th}}=0$.}\label{fig3}
\end{figure}

\begin{figure}
	\centering
	\includegraphics[width=1.0\linewidth]{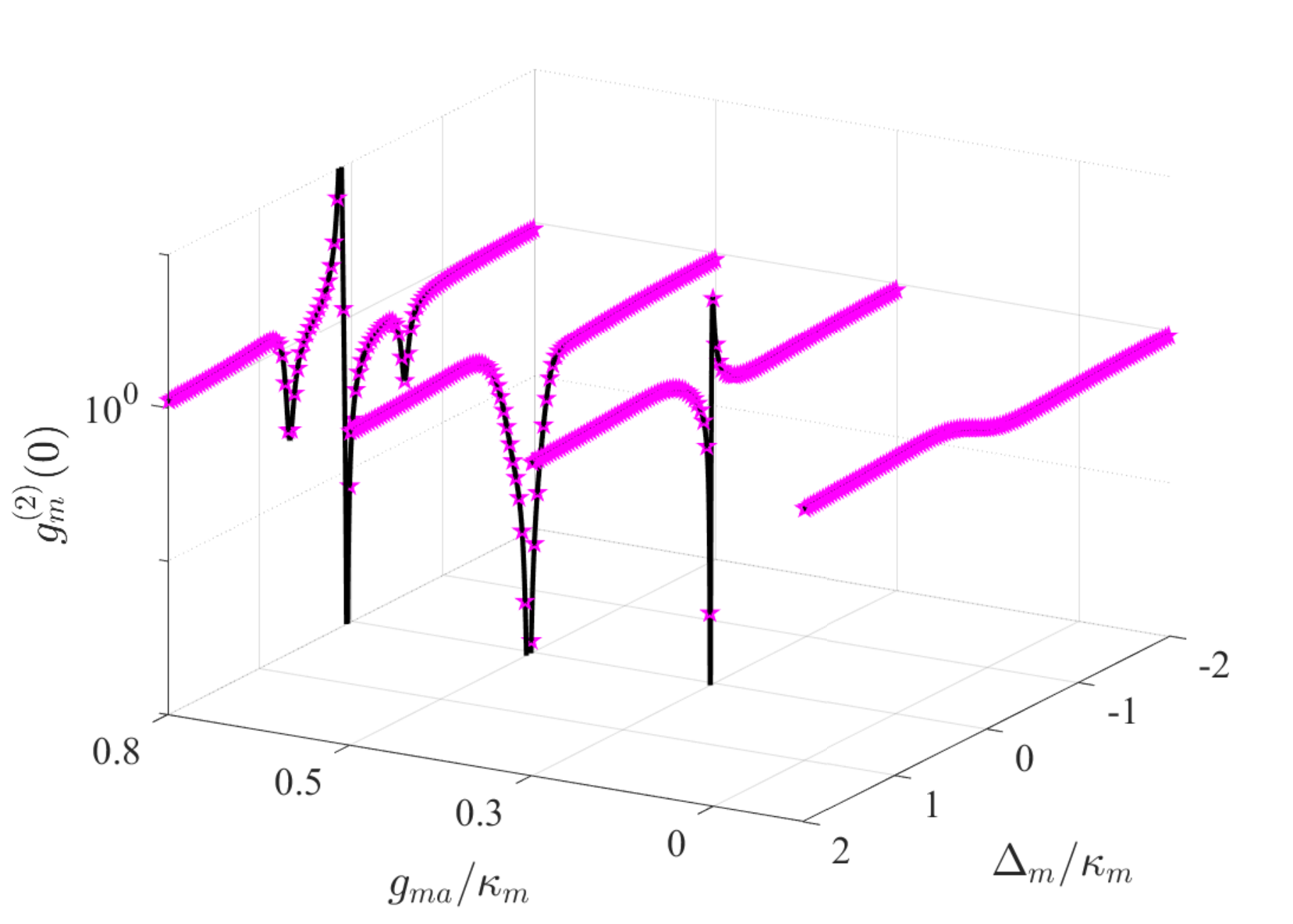}	
	\caption{(Color online) The second-order correlation $g_m^{(2)}(0)$ as a function of the detuning $\Delta_m$. The black solid line represents numerical solution and the pink star line represents the analytical solution, respectively. The values of $g_{ma}/\kappa_m$ are considered as $g_{ma}/\kappa_m=0$, $g_{ma}/\kappa_m=0.3$, $g_{ma}/\kappa_m=0.5$, $g_{ma}/\kappa_m=0.8$, the other parameters are the same as in Figure 3.
	}\label{fig4}
\end{figure}

To prove the effectiveness of the above calculations, we plot the second-order correlation function $g_m^{(2)}(0)$ versus the detuning $\Delta_m$ in \textbf{Figure~\ref{fig3}}, in which $g_m^{(2)}(0)$ is analytically calculated by Schr\"{o}dinger equation Equation (10) and numerically calculated by quantum master equation Equation (11), respectively. Apparently, the numerical result subscribes to the analytical result under the weak driving condition. In order to explore the magnon blockade effect in magnomechanical system, we plot the second-order correction function $g_m^{(2)}(0)$ versus the detuning $\Delta_{m}$ under different coupling strength regimes, as shown in Fig.~\ref{fig4}. The numerical results reveal that, when the coupling strength satisfies $g_{ma}/\kappa_m=0$, the system can be reduced to a double-mode magnetomechanical system with the absence of the cavity mode, which means that the magnon blockade cannot be achieved in a perfect way (the curve does not have the sharp dip). With the coupling strength increasing to $g_{ma}=0.5\kappa_{m}$ gradually, we find that the perfect magnon blockade effect occurs at the optimal detuning (see the curve of $g_{ma}=0.3\kappa_{m}$ in \textbf{Figure~\ref{fig4}}. We stress that this kind of perfect magnon blockade effect only possesses one dip and the width of the dip enlarges with the increase of $g_{ma}$. The reason of the only one dip is that the magnetomechanical system is under the broken $\mathcal{PT}$-symmetry region when the coupling strength satisfies $g_{ma}\in[0,~0.5\kappa_{m})$. Especially, when the coupling strength reaches $g_{ma}=0.5\kappa_{m}$, the magnetomechanical system corresponds to a critical point of the phase transition. More concretely, the magnetomechanical system is under the unbroken $\mathcal{PT}$-symmetry region when the coupling strength takes $g_{ma}>0.5\kappa_{m}$. The unbroken $\mathcal{PT}$-symmetry leads the appearance of the two new dips on both sides of optimal detuning, indicating that the system has three dips in the unbroken $\mathcal{PT}$-symmetry region.
\begin{figure*}
\centering
\includegraphics[width=0.3\linewidth]{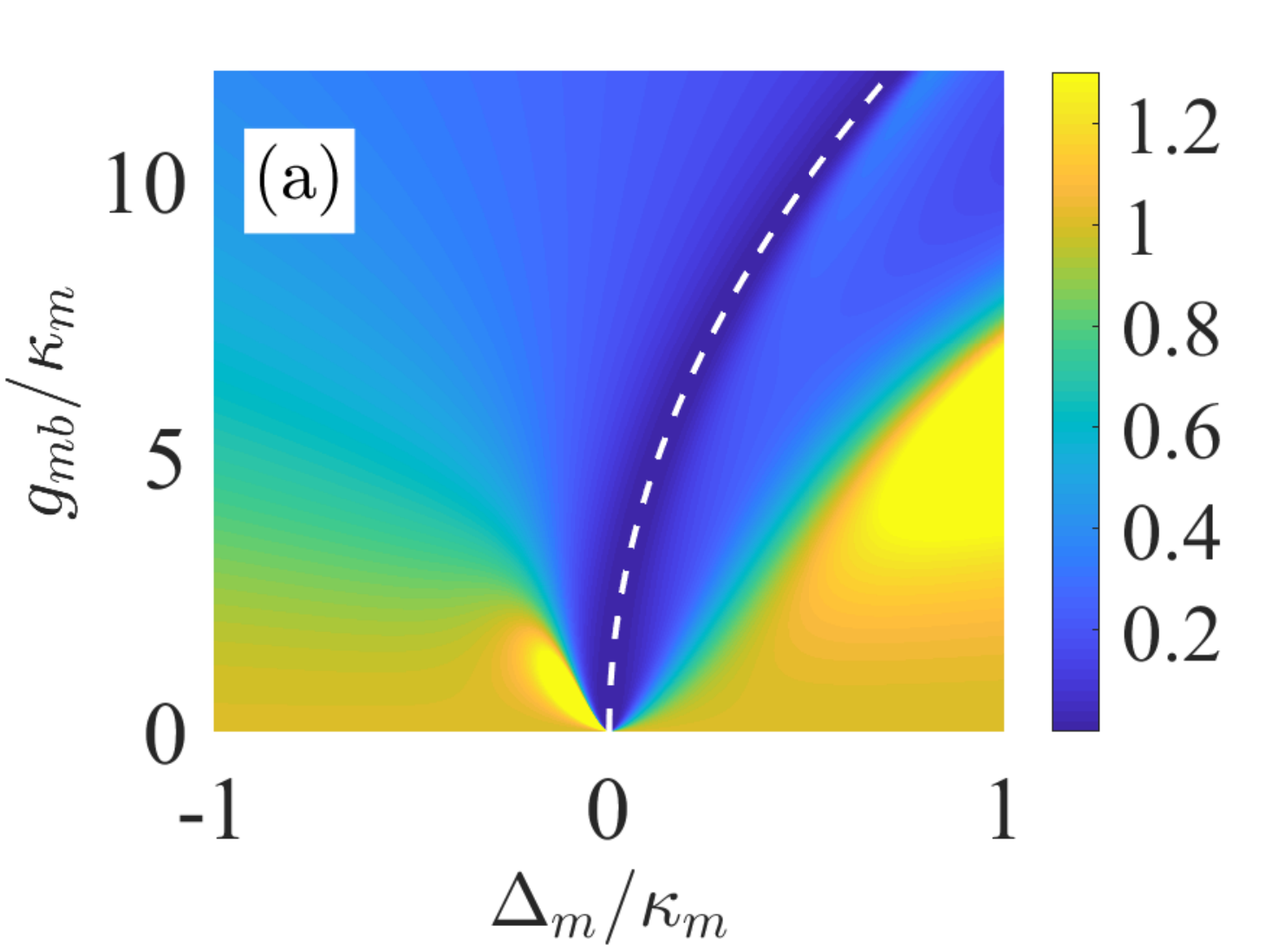}
\hspace{0.0in}
\includegraphics[width=0.3\linewidth]{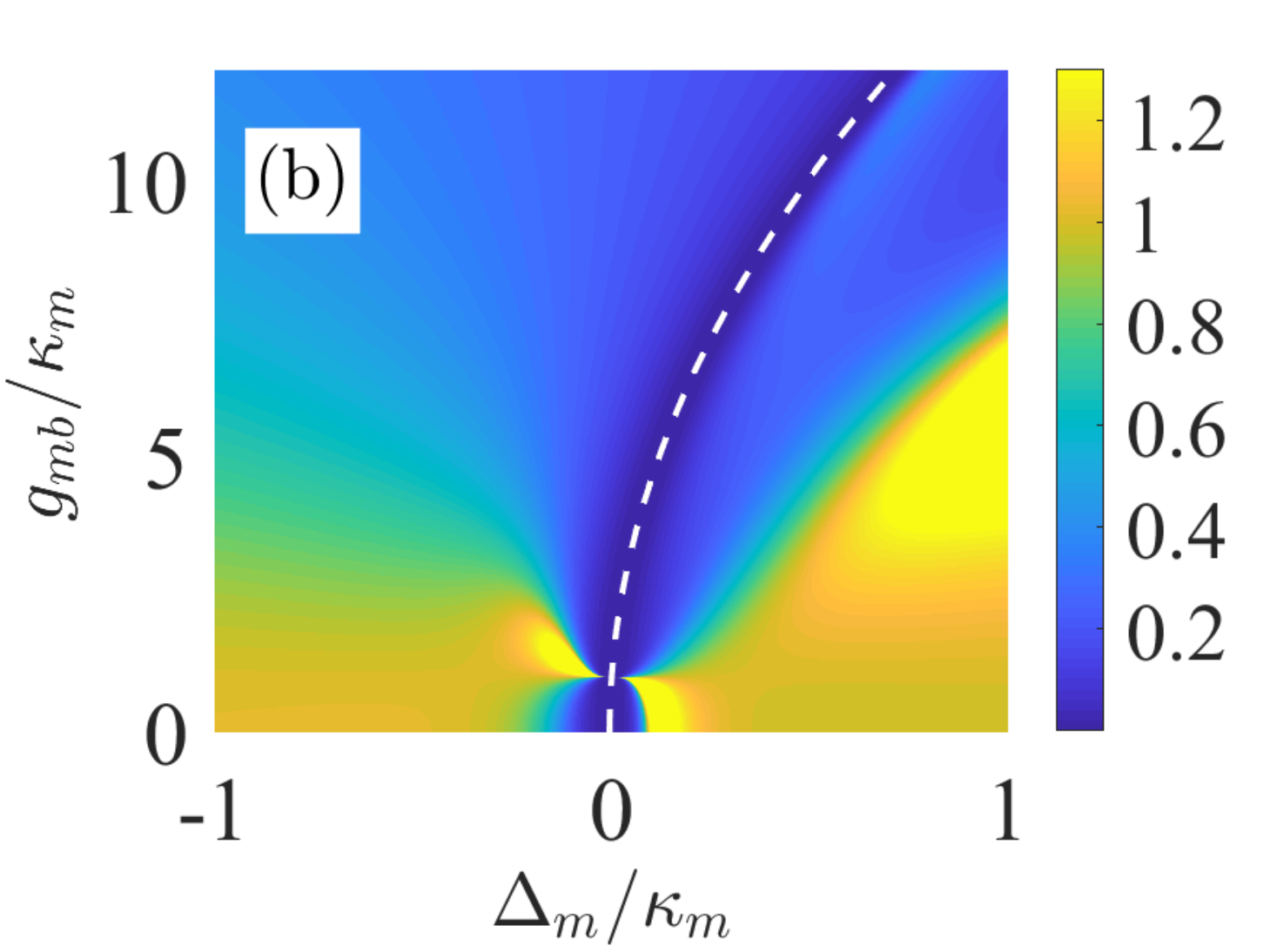}
\hspace{0.0in}
\includegraphics[width=0.3\linewidth]{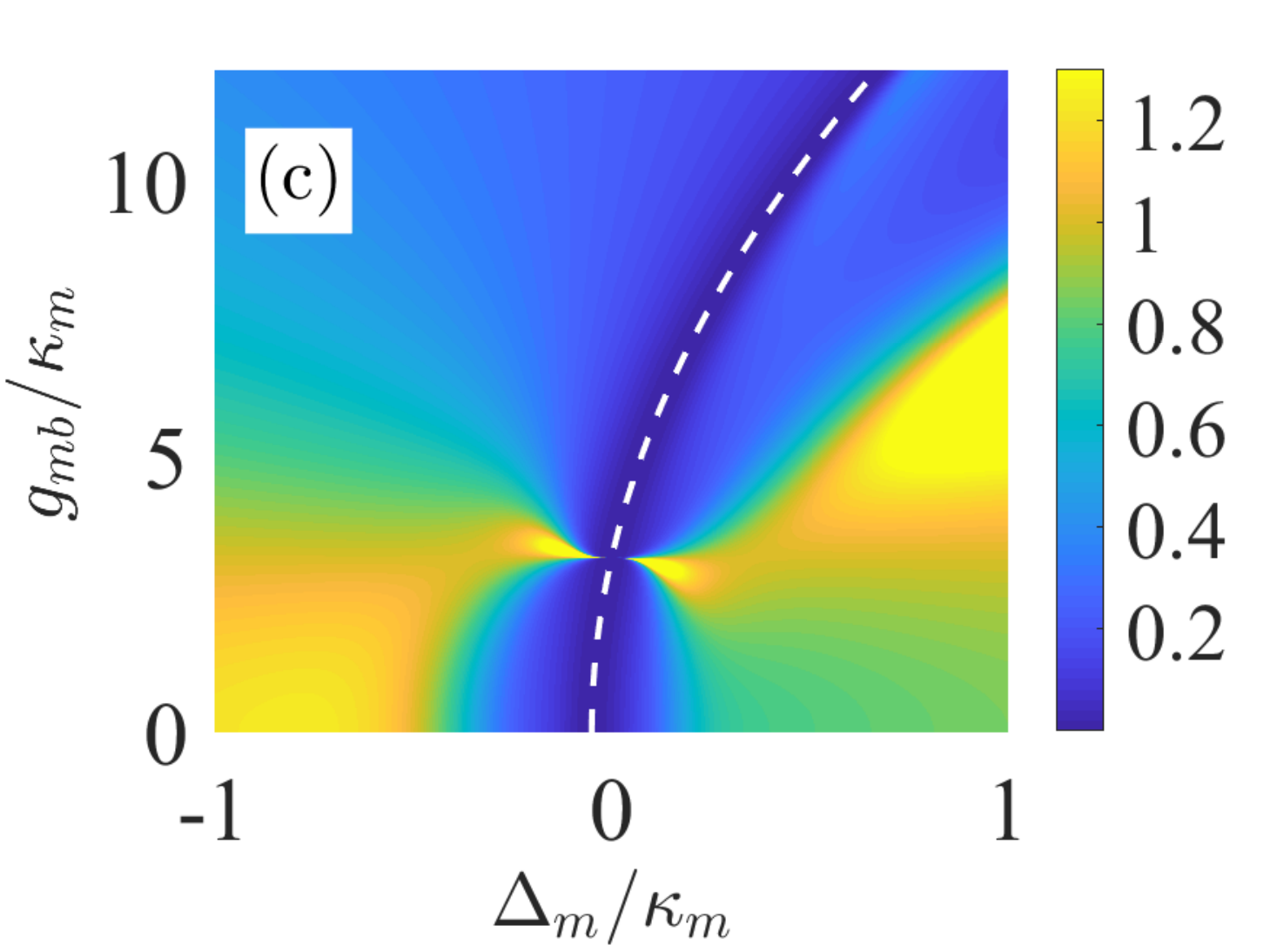}
\caption{(Color online) The second-order correlation function $g_m^{(2)}(0)$ as a function of $\Delta_m/\kappa_m$ and $g_{mb}/\kappa_m$. For different values of $K/\kappa_m$ are set as (a) $K/\kappa_m=0$, (b) $K/\kappa_m=0.01$, (c) $K/\kappa_m=0.1$, and the other parameters are the same as in Figure 3.}\label{fig5}
\end{figure*}
\begin{figure*}
	\centering
	\includegraphics[width=0.3\linewidth]{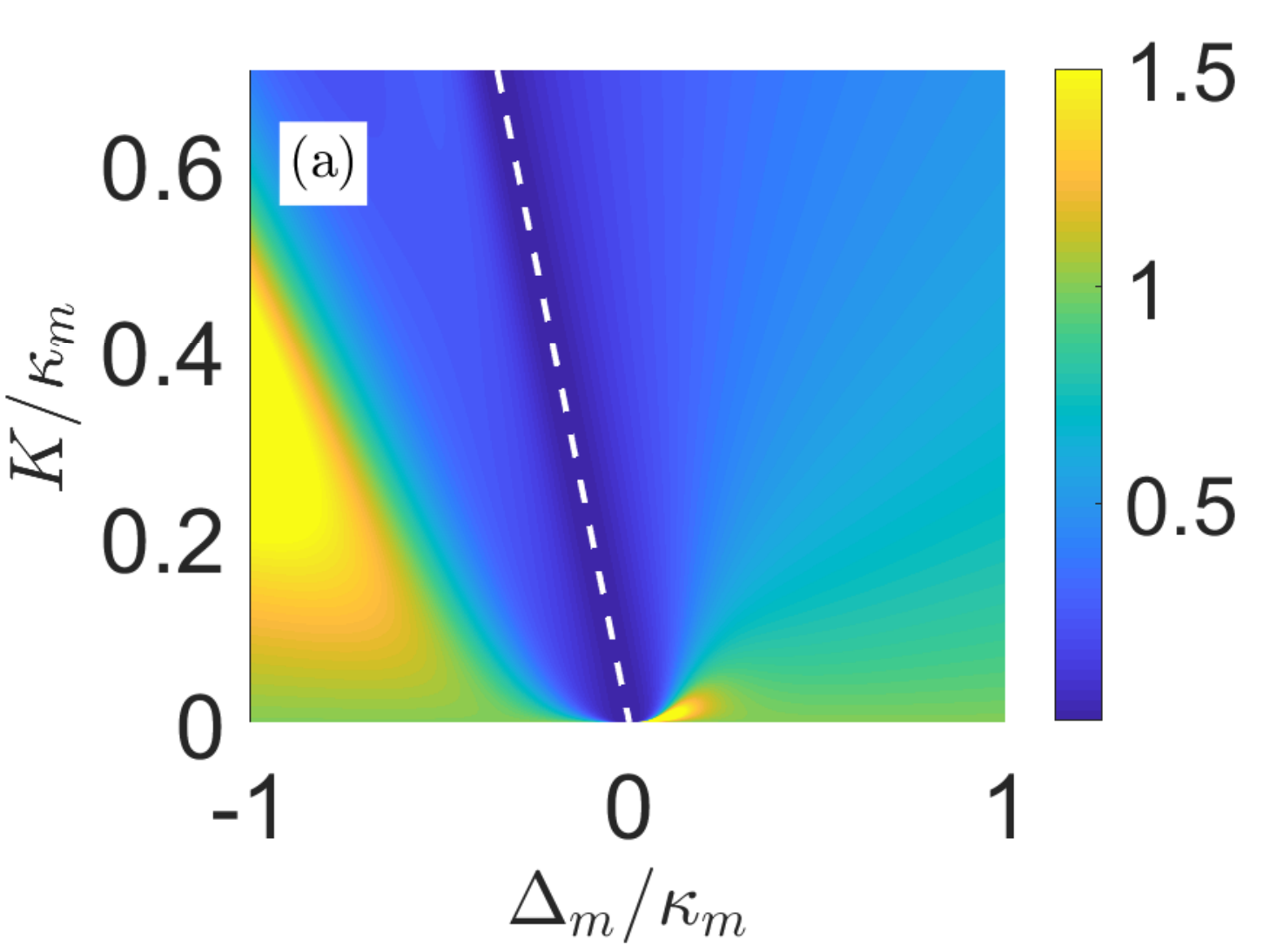}
	\hspace{0.0in}
	\includegraphics[width=0.3\linewidth]{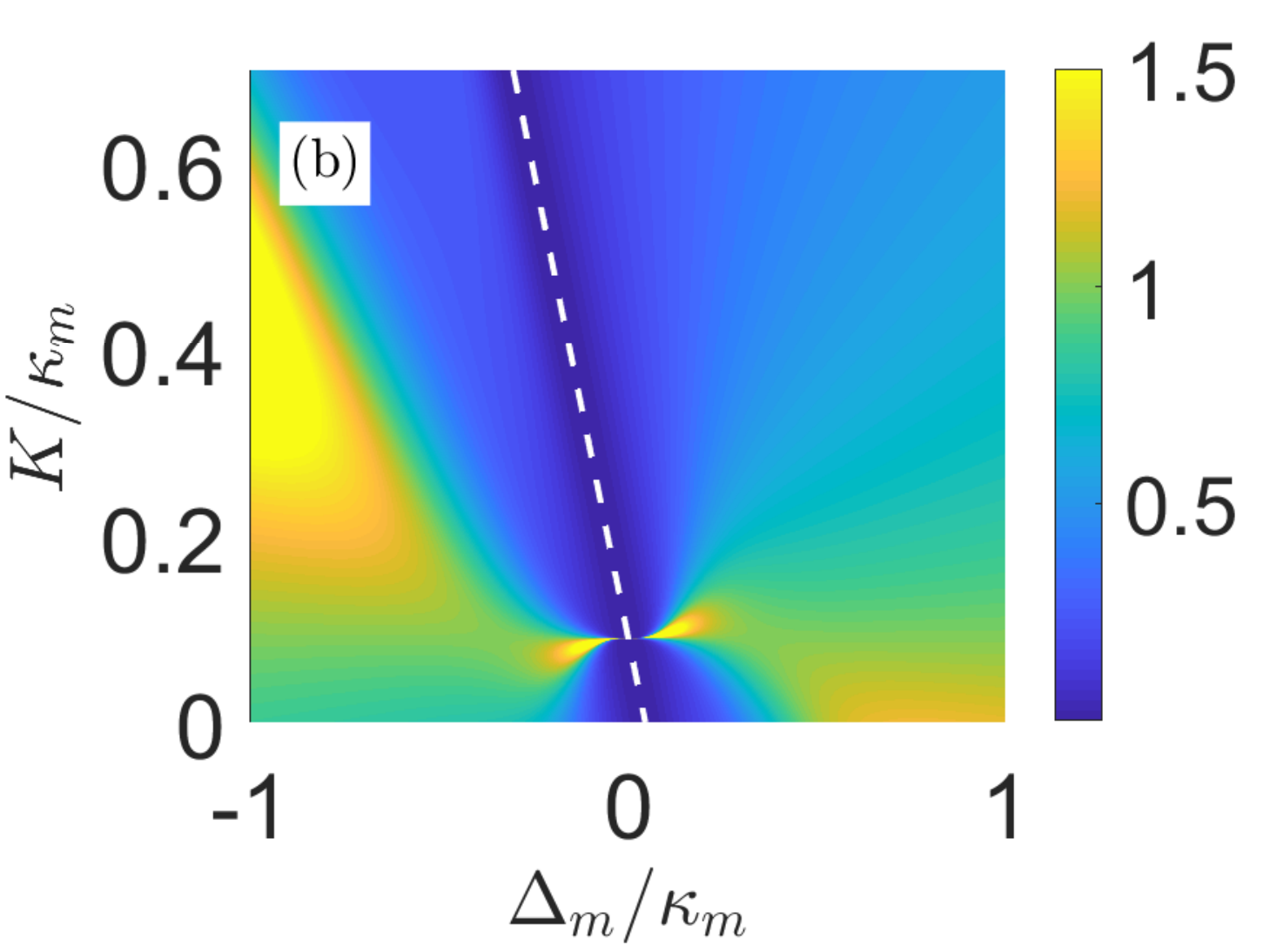}
	\hspace{0.0in}
	\includegraphics[width=0.3\linewidth]{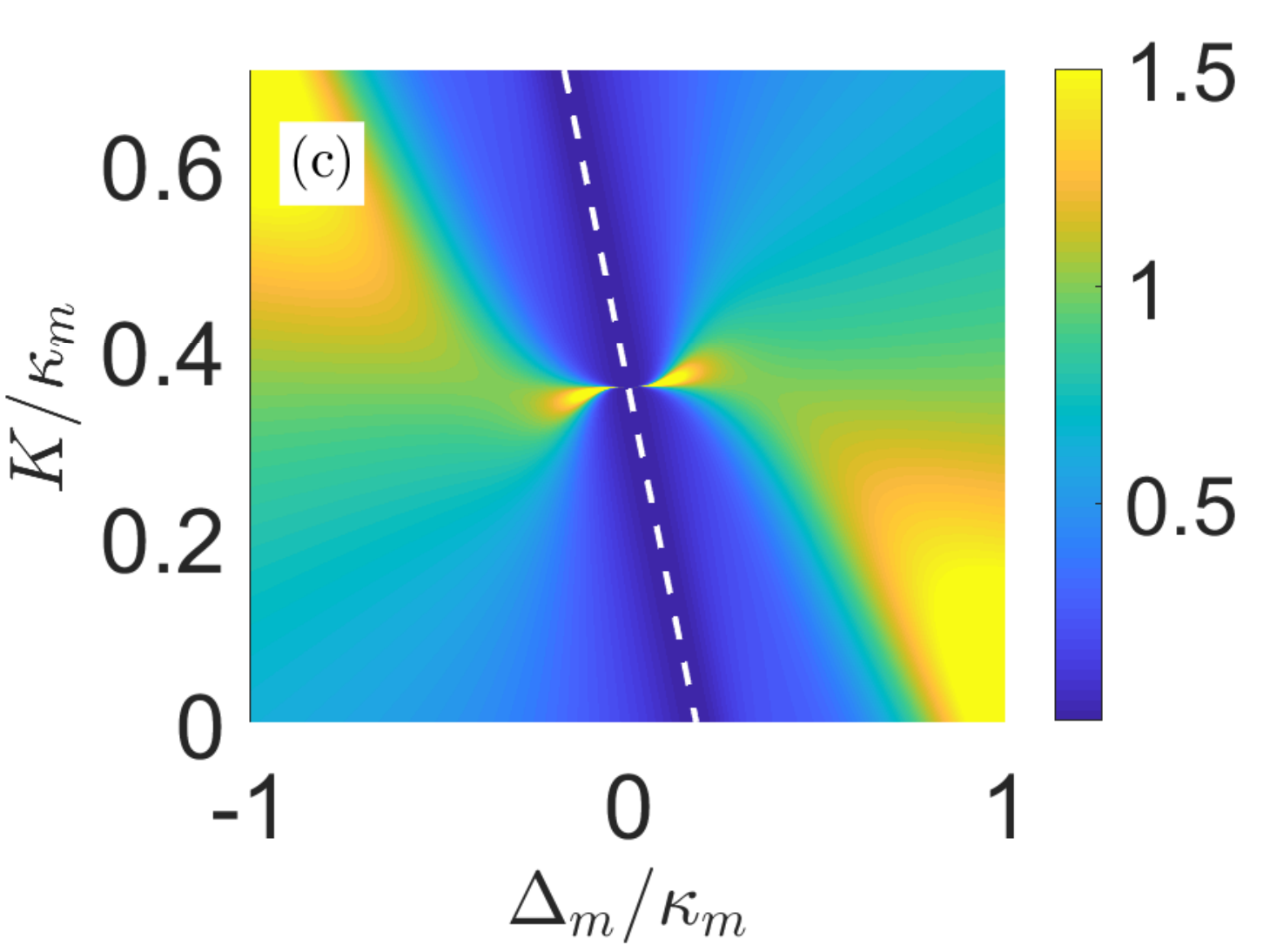}
	\caption{(Color online) The second-order correlation function $g_m^{(2)}(0)$ as a function of $\Delta_m/\kappa_m$ and $K/\kappa_m$. For different values of $g_{ma}/\kappa_m$ considered as (a) $g_{ma}/\kappa_m=0$, (b) $g_{ma}/\kappa_m=3$, (c) $g_{ma}/\kappa_m=6$, and the other parameters are the same as in Figure 3.}\label{fig6}
\end{figure*}

To study the physical mechanism of the conventional magnon blockade, we also analyze the energy of system and plot the energy level diagram in \textbf{Figure~\ref{fig1}}c.  Here, the magnon blockade indicates that the magnon cannot excite two or more magnons at the same time, but can only be excited separately. The single-magnon resonance transition is driven by external driving magnetic field and the second-magnon off-resonance transition is due to the anharmonicity of the eigenenergy spectrum, leading to that the optimal magnon blockade occurs. In order to see more clearly, we also plot the state probabilities as a function of the detuning $\Delta_m$ with two different excited number probability amplitude in Appendix. The locations of the peaks correspond to the distributed probabilities of state.

We plot the second-order correlation function $g_m^{(2)}(0)$ as a function of $\Delta_m/\kappa_m$ and $g_{mb}/\kappa_m$ in \textbf{Figure~\ref{fig5}}. We can see that the perfect magnon blockade occurs at the optimal value (see dashed white line), which comes from the analytical calculations. In the absence of Kerr nonlinear $K$, as shown in figure 5(a), the location of the perfect blockade occurring nonlinearly shift to the right with the increase of coupling strength $g_{mb}$. The phenomena of magnon blockade effect in figs.~\ref{fig5}(b)-~\ref{fig5}(c) respectively corresponds to the cases of $K/\kappa_m=0.01$ and $K/\kappa_m=0.1$, which means that the Kerr nonlinear effect exists in the system. We find that the Kerr nonlinearity does not change the overall of the perfect magnon blockade. Instead, it just moves the figure up in the parameter space. That is because the Kerr nonlinear linearly shifts the optimal location of the perfect magnon blockade occurring, which can be seen from the analytical calculation in Eq.~(\ref{eq16}). We also plot $g_m^{(2)}(0)$ as a function of $\Delta_m/\kappa_m$ and $K/\kappa_m$ in \textbf{Figure~\ref{fig6}}, in which the values of $g_{ma}/\kappa_m$ are set as $g_{ma}/\kappa_m=0$, $g_{ma}/\kappa_m=3$, and $g_{ma}/\kappa_m=6$ in figs.~\ref{fig6}(b)-~\ref{fig6}(c). To all appearances, the perfect magnon blockade effect can occur regardless of magnon-photon interaction $g_{ma}$. Meanwhile, the location of the white dashed line moves to the right all the way.

\section{\label{sec.5}Conclusions}
In conclusion, we have investigated the magnon blockade effect in a $\mathcal{PT}$-symmetric-like cavity magnomechanical system, which consists of microwave cavity mode, magnon mode, and phonon mode. Based on the broken and unbroken $\mathcal{PT}$-symmetric regions, we study the phase transition behavior. By solving the second-order correlation fuction analytically and numerically from the points of Schr\"{o}dinger equation and master equation, respectively, we derive the optimal value of detuning for the perfect magnon blockade. Moreover, we find that the magnon blockade effect can be achieved by either adjusting the coupling strength or the nonlinear parameter. In order to further understand the phenomenon of magnon blockade, we also discuss the different blockade mechanisms. The results indicate that one dip corresponds to the unconventional magnon blockade while two new dips to the conventional magnon blockade. Our scheme provides a feasible method to realize the magnon blockade in the weak parameter regime and we hope that it could pave a way to realize the magnon blockade in experiment.

\section*{Acknowledgements}
This work was supported by the National Natural Science Foundation of China under Grant Nos.
61822114, 61465013, and 11465020.

\section*{Conflict of Interest}
The authors declare no conflict of interest.

\section*{Appendix}
\section*{CALCULATION OF DETUNING}

\begin{figure}
\centering
\includegraphics[width=0.48\linewidth]{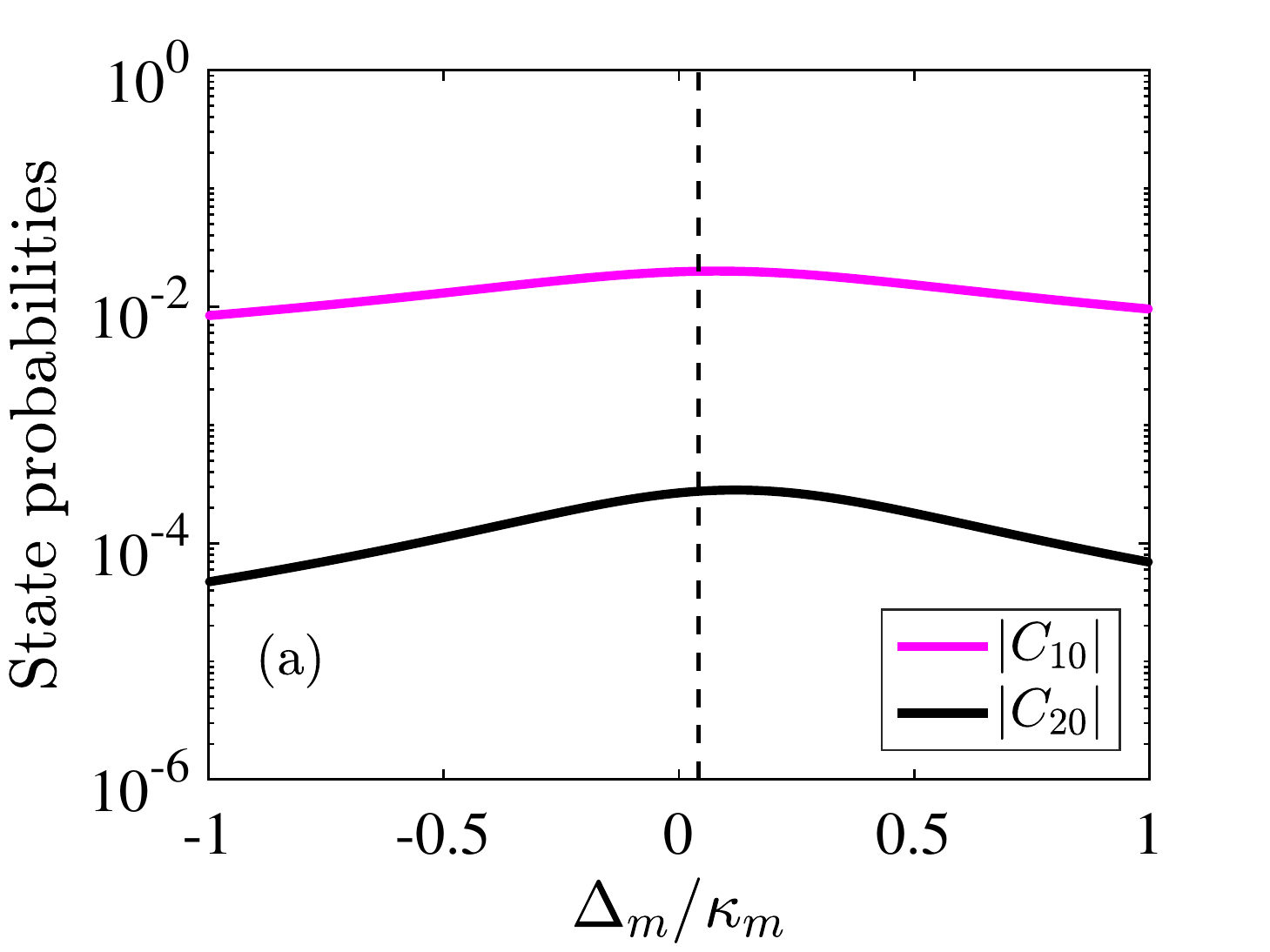}
\hspace{0.0in}
\includegraphics[width=0.48\linewidth]{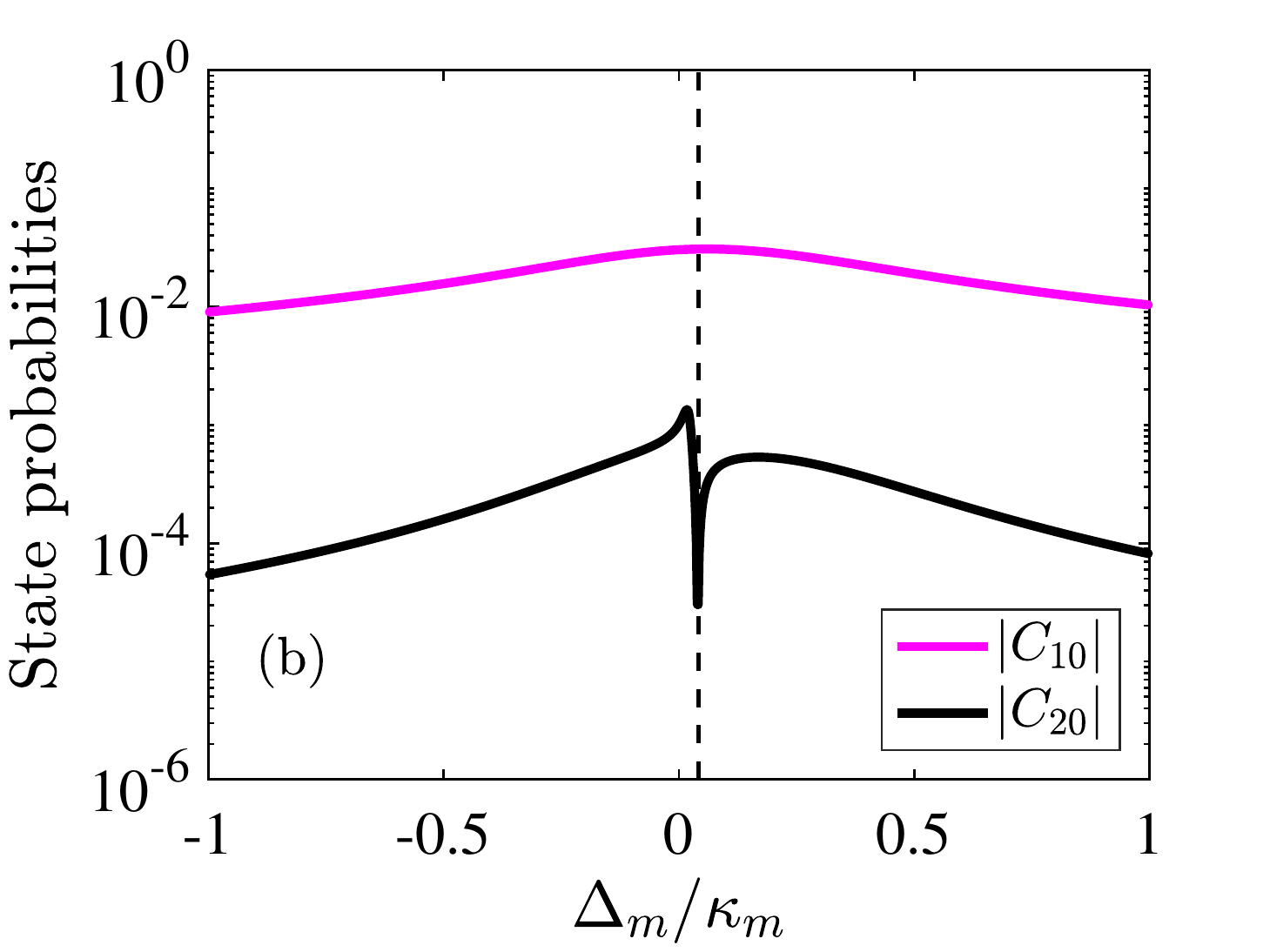}
\hspace{0.0in}
\includegraphics[width=0.48\linewidth]{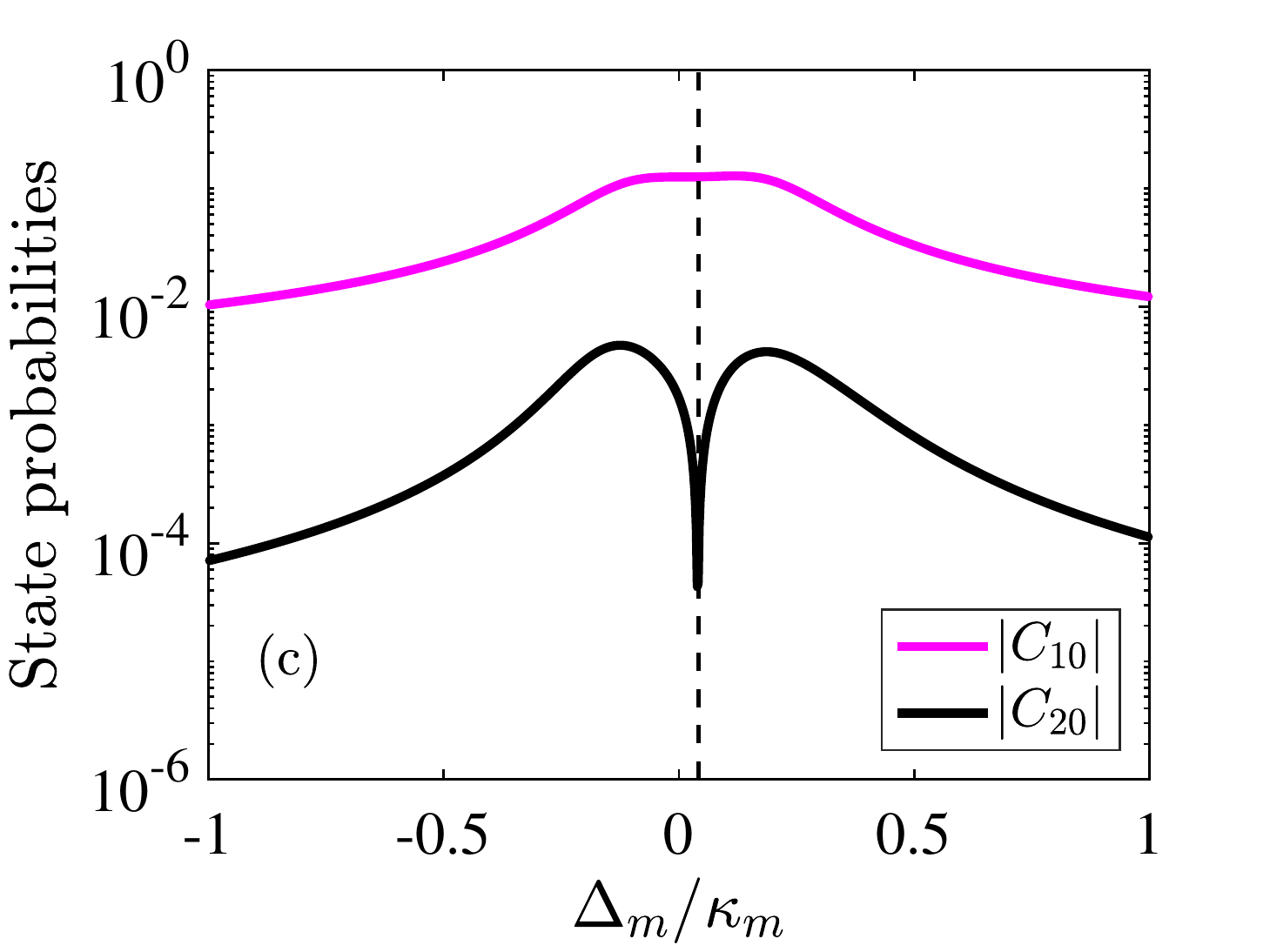}
\hspace{0.0in}
\includegraphics[width=0.48\linewidth]{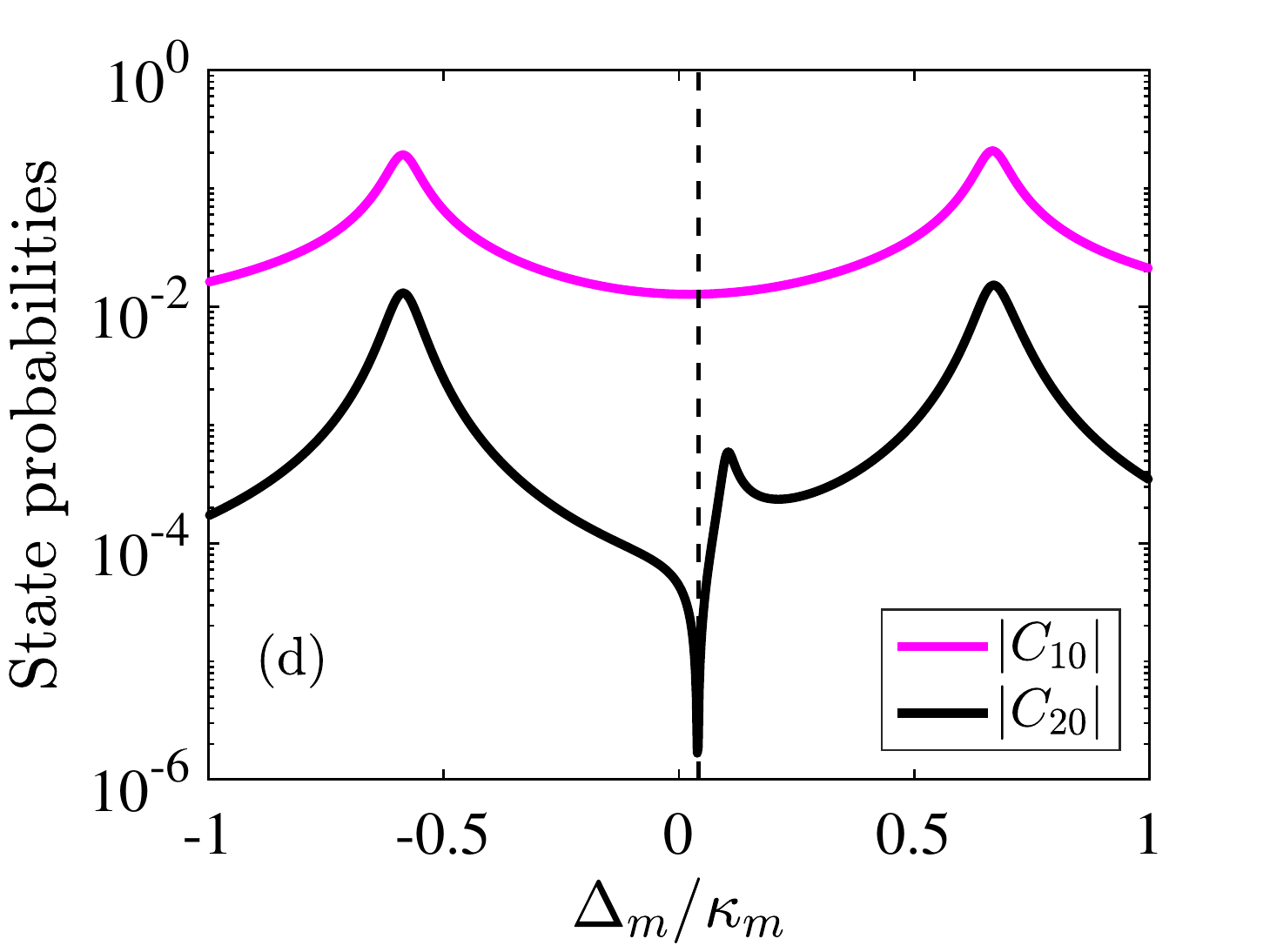}	
\caption{(Color online) The State probabilities as a function of the detuing $\Delta_m$ with two different probablity amplitudes. The pink curves indicate probablity amplitude $\vert C_{10} \vert$, the black curves indicate probablity amplitude $\vert C_{20} \vert$. The system parameters are set as (a) $g_{ma}/\kappa_m=0$, (b) $g_{ma}/\kappa_m=0.3$, (c) $g_{ma}/\kappa_m=0.5$, (d) $g_{ma}/\kappa_m=0.8$, the other experimental parameters are the same as in Figure 3.
	}\label{fig7}
\end{figure}

In the steady-state, we study the magnon blockade in the magnonmechanical system, it exists $\{{C_{02}, C_{11}, C_{20}}\}\ll\{{C_{01}, C_{10}}\}\ll\ C_{00}$, so we obtain linear equations about the expanding coeffcients as follows,
\begin{eqnarray}\label{eq14}
C_{10}&=&\frac{\Omega(\Delta_m+i\kappa_m/2)}{g_{ma}^2-(\Delta_m+i\kappa_m/2)(\Delta_m-i\kappa_m/2+K-g_{mb}^2/\omega_b)},\cr\cr	C_{01}&=&\frac{-\Omega g_{ma}}{g_{ma}^2-(\Delta_m+i\kappa_m/2)(\Delta_m-i\kappa_m/2+K-g_{mb}^2/\omega_b)},\cr\cr	C_{11}&=&\frac{16\Omega^2\omega_b^2g_{ma}(2\Delta_m+i\kappa_m)[\omega_b(\Delta_m+K)-g_{mb}^2]}{M},\cr\cr	C_{02}&=&\frac{16\sqrt2\Omega^2g_{ma}^2\omega_b^2[\omega_b(\Delta_m+K)-g_{mb}^2]}{M},\cr\cr	C_{20}&=&\frac{2\sqrt2\Omega^2\omega_b^2(2\Delta_m+i\kappa_m)^2[\omega_b(2\Delta_m+K)-g_{mb}^2]}{M},
\end{eqnarray}
with
\begin{eqnarray}\label{eq15}
M&=&g_{mb}^2\omega_b[8g_{ma}^2-(2\Delta_m+i\kappa_m)(10\Delta_m-i\kappa_m+8K)]\cr\cr&&
+\{\omega_b^2[(2\Delta_m+K)(4\Delta_m^2+\kappa_m^2+8\Delta_mK+4i\kappa_mK)\cr\cr&&
-8g_{ma}^2\Delta_m-8g_{ma}^2K]\}
\{4g_{ma}^2\omega_b+(2\Delta_m+i\kappa_m)\cr\cr&&
[2g_{mb}^2+\omega_b(i\kappa_m-2\Delta_m-2K)]\}\cr\cr&&
+8g_{mb}^4\Delta_m+4ig_{mb}^4\kappa_m.
\end{eqnarray}
To achieve the optimal parameters, we can set $|C_{20}|=0$, which satisfies the condition
\begin{eqnarray}\label{eq16}
K=-2\Delta_m.
\end{eqnarray}

We plot the probability amplitude in four different coupling strength regimes in \textbf{Figure~\ref{fig7}}. The results reveal that, when the coupling strength satisfies $g_{ma}=0$, the effect of detuning $\Delta_m$ on the state probabilities of $|C_{10}|$ and $|C_{20}|$ is tiny, as shown in fig.~\ref{fig7}(a). With the coupling strength increasing to $g_{ma}=0.3\kappa_{m}$ gradually, there is only one dip in the vicinity of the optimal detuning for $|C_{20}|$ (see the black line). Nevertheless, the effect of $g_{ma}$ on the system is tiny for $|C_{10}|$ (see the pink line) in fig.~\ref{fig7}(b). Especially, when the coupling strength reaches $g_{ma}=0.5\kappa_{m}$, the magnetomechanical system corresponds to a critical point of the phase transition in fig.~\ref{fig7}(c). The reason of the one dip is that the magnetomechanical system is under the broken-$\mathcal{PT}$-symmetric region when the coupling strength satisfies $g_{ma}\in[0,~0.5\kappa_{m})$. Specifically, we find that the pink line has a split and forms two new symmetrical peaks in the vicinity of optimal detuning in fig.~\ref{fig7}(d), which belongs to unbroken-$\mathcal{PT}$-symmetry, and the distance between the two peaks enlarge with further increasing $g_{ma}$. One can note that the magnon blockade effect is different in this system. We can find that there is only one dip when $g_{ma}\in[0,~0.5\kappa_{m})$, and the one dip corresponds to the locations of the conventional magnon blockade. While $g_{ma}>0.5\kappa_{m}$, the two new dips correspond to the locations of the conventional magnon blockade.



\end{document}